\documentclass{appolb}

\usepackage{graphicx}
\usepackage{epstopdf}
\usepackage{amsmath,amssymb}
\usepackage{multirow}

\newcommand{\beq}{\begin{equation}}
\newcommand{\eeq}{\end{equation}}
\newcommand{\bea}{\begin{eqnarray}}
\newcommand{\eea}{\end{eqnarray}}
\newcommand{\nn}{\nonumber}
\newcommand{\eq}{Eq.~}

\newcommand{\fig}{Fig.~}

\newcommand{\tr}{{\rm Tr}}
\newcommand{\bx}{{\bf x}}
\newcommand{\by}{{\bf y}}

\def\lsi{\raise0.3ex\hbox{$<$\kern-0.75em\raise-1.1ex\hbox{$\sim$}}}
\def\gsi{\raise0.3ex\hbox{$>$\kern-0.75em\raise-1.1ex\hbox{$\sim$}}}

\begin{document}
\title{QCD in the heavy dense regime:\\ Large $N_c$ and quarkyonic matter%
\thanks{Presented at `Criticality in QCD and the hadron resonance gas', Wroclaw, July 29-31, 2020}%
}
\author{Owe Philipsen\footnote{Speaker}, Jonas Scheunert
\address{ITP, Goethe-Universit\"at Frankfurt\\
Max-von-Laue-Str.1\\
60438 Frankfurt am Main, Germany}
}

\maketitle
\begin{abstract}
After combined character and hopping expansions and integration over the spatial gauge links, 
lattice QCD reduces to a three-dimensional $SU(3)$ Polyakov loop model with complicated interactions.
A simple truncation of the effective theory is valid for heavy quarks on reasonably fine lattices and can be solved by
linked cluster expansion in its effective couplings. This was used ealier to demonstrate the onset transition
to baryon matter in the cold and dense regime. Repeating these studies for general $N_c$, one finds
that for large $N_c$ the onset transition becomes first-order, and the pressure scales as $p\sim N_c$ 
through three consecutive orders in the hoppoing expansion. 
These features are consistent with the formal definition of quarkyonic matter given
in the literature. We discuss the implications for $N_c=3$ and physical QCD. 
 
\end{abstract}
  
\section{Introduction}

The physics of cold and dense baryon matter is of ever increasing interest in view of growing observational data
from neutron stars and their mergers, as well as heavy-ion collisions at low energies and large densities.
To date, the corresponding parametric regime of QCD is inaccessible to lattice simulations because of a severe
sign problem for baryo-chemical potential $\mu_B\neq 0$. The low density regime $\mu_B < 3T$, where the crossover from a
hadron resonance gas to a quark gluon plasma takes place, can be controlled by a number of methods, 
and no sign of criticality is observed \cite{review1,review2}. Here we address the cold and dense regime of the nuclear
liquid gas transition, corresponding to large $\mu_B/T$, within an effective lattice theory derived analytically from full QCD.
While this theory is only valid for heavy quarks, the qualitative features of the nuclear liquid gas transition
can be reproduced directly from QCD in this framework. In particular, we discuss here the possibility for quarkyonic
matter as conjectured in \cite{quarky}.   

\section{An effective lattice theory for heavy quarks}

%
%
We start with lattice QCD in the Wilson fomulation
at finite temperature, i.e.,  with compact euclidean time dimension of
$N_\tau$ slices, $T=1/(aN_\tau)$, and (anti-)periodic boundary conditions for (fermions) bosons. 
An effective theory in terms of temporal links only
is obtained after integrating over the quark fields and gauge links in spatial directions in the partition function,
\beq
Z=\int DU_0DU_i\;\det Q \; e^{-S_g[U]}\equiv\int DU_0\;e^{-S_{eff}[U_0]}=\int DW \,\;e^{-S_{eff}[W]}\;.
\eeq 
The effective action then depends on temporal Wilson lines $W({\bf x})$ closing through 
the periodic boundary, or Polyakov loops $ L({\bf x})=\tr W({\bf x})$, and is in principle
unique and exact. 
In practice, we first expand the QCD action in powers of the coefficient of the fundamental character $u$
and the hopping parameter $\kappa$,
\beq
u(\beta)=\frac{\beta}{18}+\frac{\beta^2}{216}+\ldots < 1, \qquad \kappa=\frac{1}{2am_q+8}\;.
\eeq
The dependence of $u$ on the lattice gauge coupling $\beta=2N_c/g^2$ is known to arbitrary precision, and
$u$ is always smaller than one for finite $\beta$-values. 
Since the hopping expansion is in inverse quark mass, the effective theory to low orders is valid for heavy quarks only.
Both expansions result in convergent series within a finite radius of convergence.
Truncating these at some finite order, the integration over the
spatial gauge links can be performed analytically to provide a closed expression for the effective theory.
The integration over spatial links causes long-range interactions 
of Polyakov loops at all distances and to all powers, which must be taken into account according
to the power counting of the expansion parameters. The first terms of the partition function, with nearest neighbour interactions only, then read \cite{efft1,fromm}
\bea
\label{zpt}
Z&=&\int DW\prod_{<\bx, \by>}\left[1+\lambda(L_{\bx}L_{\by}^*+L_{\bx}^*L_{\by})\right]\\
&& \hspace*{-0.5cm}
\times \prod_{\bx}[1+h_1L_{\bx}+h_1^2L_{\bx}^*+h_1^3]^{2N_f}[1+\bar{h}_1L^*_{\bx}+\bar{h}_1^2L_{\bx}+\bar{h}_1^3]^{2N_f}
\prod_{<\bx, \by>}\Bigg(1-   \nn\\
&& \hspace*{-0.5cm}
2h_{2}\left({\rm Tr} \frac{h_1W_{\bx}}{1+h_1W_{\bx}} - {\rm Tr} \frac{\bar{h}_1W^\dag_{\bx}}{1+\bar{h}_1W_{\bx}}\right) 
 \left({\rm Tr} \frac{h_1W_{\by}}{1+h_1W_{\by}} - {\rm Tr} \frac{\bar{h}_1W^\dag_\by}{1+\bar{h}_1W^\dag_{\by}}\right)\Bigg)
\;.\nn
\eea
The first line corresponds to the Yang-Mills part, the second line to the static determinant, and the third line to the
leading corrections from quark hops representing pion exchange. The effective couplings are functions of the original QCD parameters (for complete expressions beyond LO, see \cite{fromm})
\bea
\lambda&=& u^{N_\tau}\exp[N_\tau(4u^4+\ldots)]\;,\nn\\
h_1&=&(2\kappa e^{a\mu})^{N_\tau}\left(1+\ldots \right) =e^{\frac{\mu-m}{T}}(1+\ldots), \quad 
\bar{h}_1=h_1(-\mu)\;,\nn\\
h_2&=&\kappa^2 N_\tau/N_c(1+\ldots)\;.
\eea
with $am=-\ln(2\kappa)=am_B/3$ the leading-order constituent quark mass in a baryon. 

This effective theoy has a mild sign problem only and can be simulated with 
reweighting or complex Langevin methods \cite{fromm,bind}.
Moreover, it  can be treated by linked-cluster expansion methods known from statistical physics. 
These have recently been tested to high order in an $SU(3)$ spin model, allowing a quantitative
determination of the phase diagram with zero or non-zero chemical potential \cite{su3spin}.
Also in the present context, a linked cluster expansion to order $\kappa^8 u^5$ was successfully
tested against numerical ones \cite{k8}. 

\subsection{The deconfinement transition}
\begin{figure}[t]
\centering
\includegraphics[width=0.45\textwidth,clip]{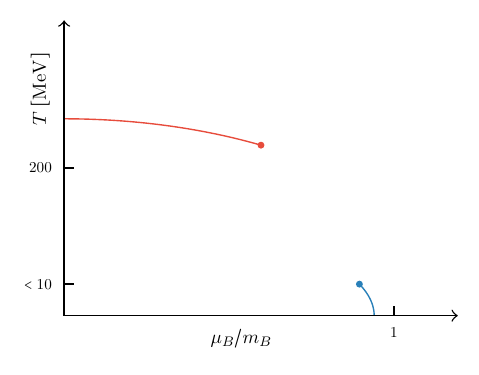}\hspace*{0.5cm}
\includegraphics[width=0.45\textwidth,clip]{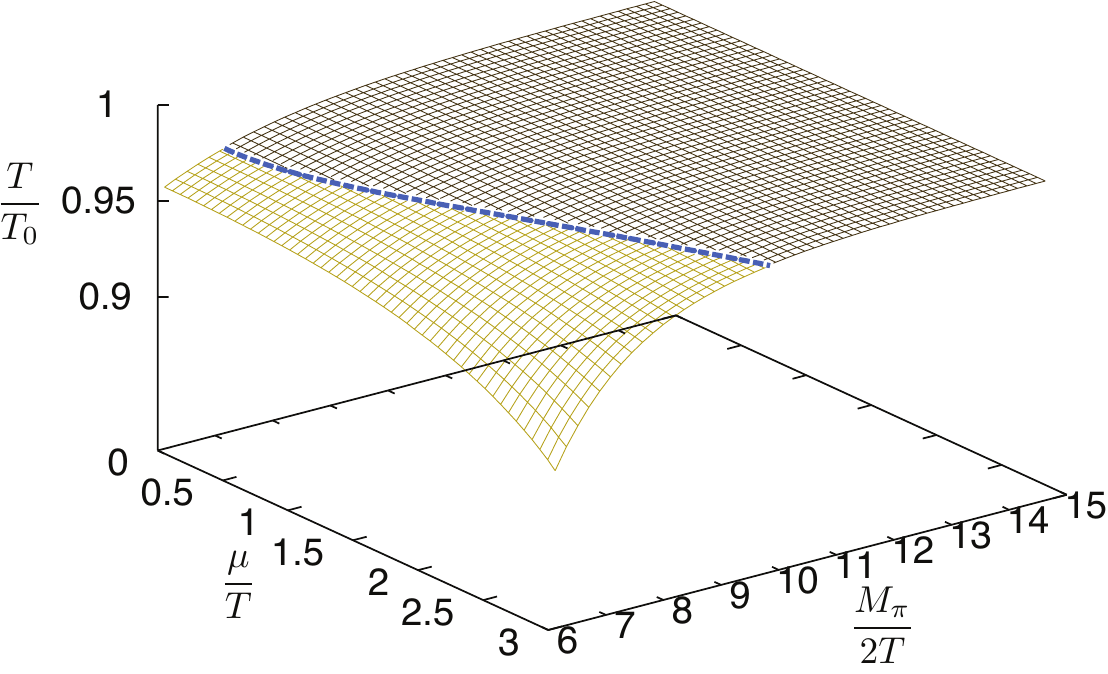}
\caption{Left: Qualitative phase diagram for QCD with very heavy quarks. Right: Deconfinement transition for $N_f=2, N_\tau=6$ 
from the 3d effective theory \cite{fromm}.}
\label{fig:heavy}       
\end{figure}

The effective theory can now be used to map out the phase diagram of QCD with heavy quarks. 
At $\mu=0$ and infinite quark mass, a first-order deconfinement transition is found at high temperature, 
corresponding to the breaking
of the $Z(3)$ center symmetry.  The phase transition weakens with decreasing quark
mass until it disappears at a critical point. This is exactly what is also found in Monte Carlo simulations of full QCD \cite{ejiri,cuteri}.
However, contrary to full QCD, the effective theory can also be applied to finite chemical potential \cite{k8}. 
For heavy quarks, the first-order deconfinement transition also weakens with chemical potential ending in 
a critical point, whose location depends on the quark mass, \fig\ref{fig:heavy}. The same qualitative behaviour is
also found by continuum Polyakov loop model studies \cite{lo,frg}.

\subsection{The baryon onset transition}

More difficult to address is the cold and dense region, since the sign problem grows exponentially with $\mu/T$.  
Here we switch to analytic methods.
It is instructive to first consider 
the strong coupling ($\beta=0$) limit with a static quark determinant. In this case the partition function 
can be solved analytically. A low temperature,
mesonic contributions are exponentially suppressed by chemical potential and for  
$N_f=1$ we have  \cite{silver, bind} 
\beq
Z(\beta=0) \stackrel{T\rightarrow 0}{\longrightarrow}z_0^V \quad \mbox{with}\quad 
z_0=1+4h_1^{3}+h_1^{6}. 
\label{eq:zsc}
\eeq
Note that this corresponds to a free baryon gas with two species. With one quark flavour only, there are no nucleons and the
first prefactor indicates a spin 3/2  quadruplet of $\Delta$'s whereas the second term is a spin 0 six quark state or di-baryon.
The quark number density is now easily evaluated
\beq
n=
\frac{T}{V}\frac{\partial}{\partial \mu}\ln Z=\frac{1}{a^3}\frac{4N_ch_1^{N_c}+2N_ch_1^{2N_c}}{1+4h_1^{N_c}+h_1^{2N_c}}\;,
 \quad \lim_{T\rightarrow 0} a^3n=\left\{\begin{array}{cc} 0, & \mu<m\\
	2N_c, & \mu>m\end{array}\right.\;.
\eeq
At zero temperature this is a step function, which
reflects the ``silver blaze'' property of QCD, i.e.~the fact that the baryon number stays zero
for small $\mu$ even though the partition function explicitly depends on it \cite{cohen}. Once the baryon chemical potential 
$\mu_B=3 \mu$
is large enough to make a baryon ($m_B=3m$ in the static strong coupling limit), a discontinuous phase transition 
to a saturated baryon crystal takes place. 
Note that saturation density here is $2N_c$ quarks per flavour and lattice
site and reflects the Pauli principle. This discretisation effect has to disappear
in the continuum limit.

In the case of two flavours the corresponding expression reads \cite{bind}
\begin{eqnarray}
z_0& =& (1 + 4 h_d^3 + h_d^6)+ (6 h_d^2 + 4 h_d^5) h_u+ (6 h_d + 10 h_d^4)h_u^2+ 
  (4 + 20 h_d^3   \\
&& + 4 h_d^6)h_u^3 + (10 h_d^2 + 6 h_d^5) h_u^4+ ( 4 h_d + 6 h_d^4) h_u^5 
  +(1 + 4 h_d^3 + h_d^6)h_u^6\;,\nn
\label{eq:freegas}  
\end{eqnarray}
where we have separate $h_1$ couplings for the $u$- and $d$-quarks. Now we identify also the 
spin 1/2 nucleons and several other baryonic multi-quark states with their correct spin degeneracy. 
Remarkably, the spin-flavour-structure
of the QCD baryons is obtained in this simple limit!

\begin{figure}[t]
\centering
\includegraphics[width=0.45\textwidth]{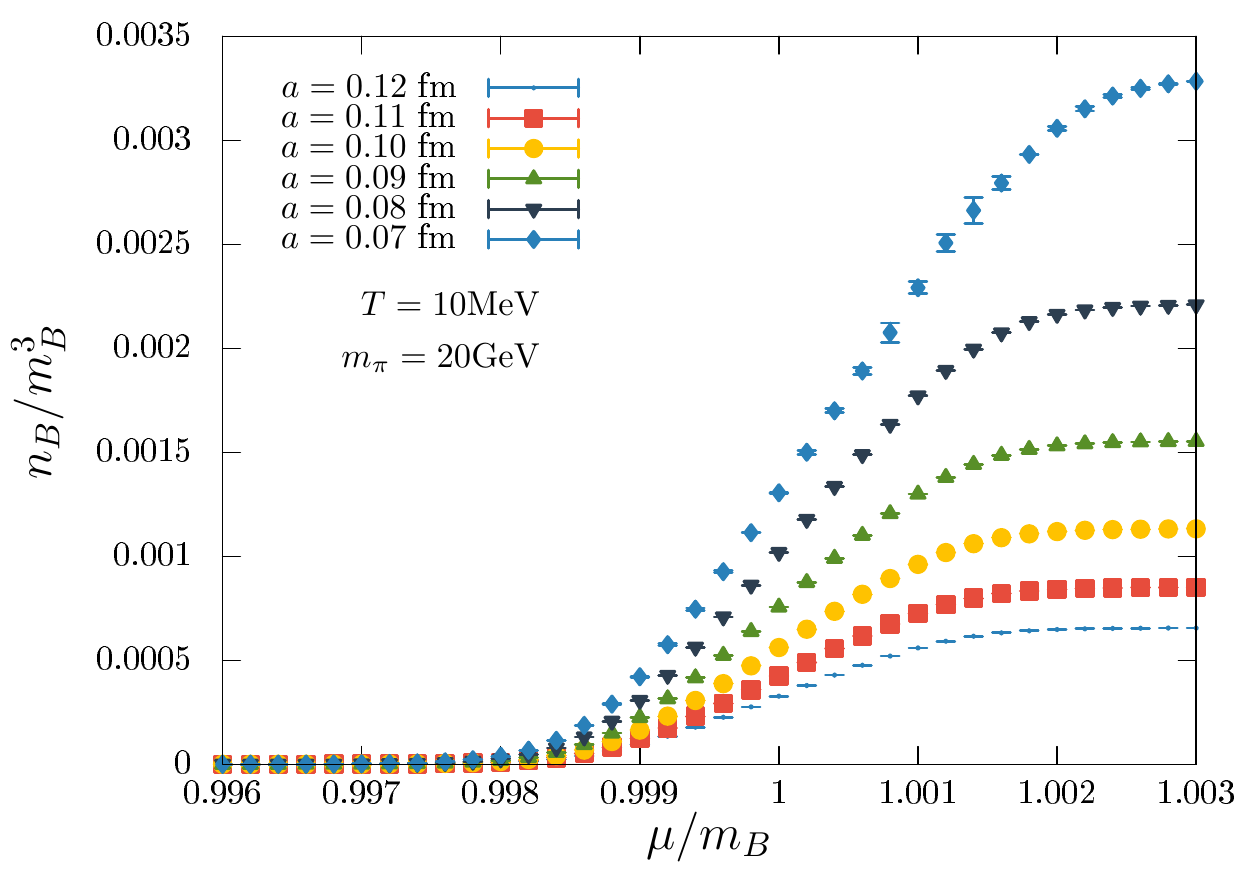}\hspace*{0.5cm}
\includegraphics[width=0.45\textwidth,clip]{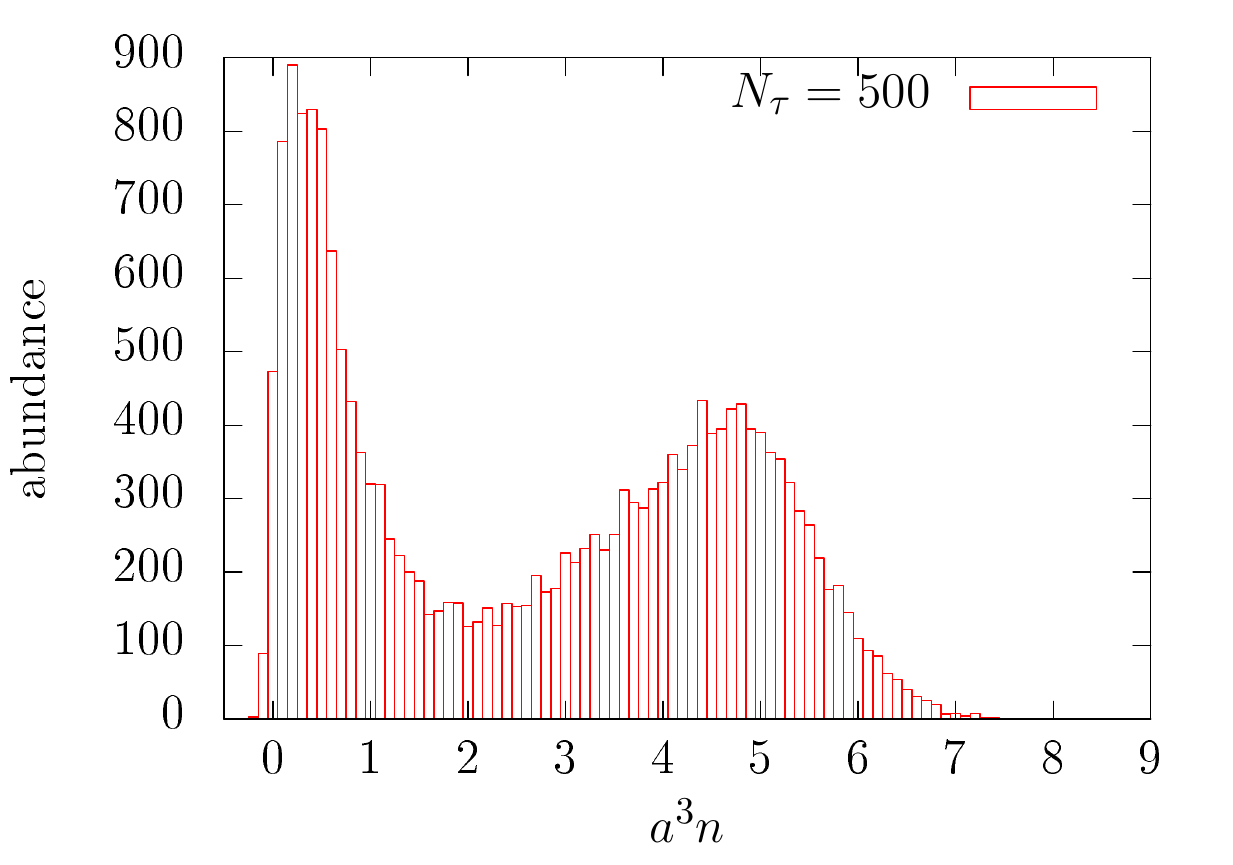}
\caption{Left: Baryon number density for heavy quarks, various lattice spacings  \cite{k8}. 
Right: Baryon number distribution at onset for low $T$ and light quarks: first-order transition. 
}
\label{fig:cont}       
\end{figure}

When corrections are added, the step function gets smoothed, as shown in \fig\ref{fig:cont} (left) in a calculation 
through orders $\kappa^8u^5$ for various lattice spacings.  As expected, the saturation level moves towards infinity as
the continuum is approached. Note however, that this makes continuum extrapolations at growing chemical potential
formidably difficult to control.
An important observation is that the onset transition happens already before $\mu_B=m_B$. This is also expected 
for the physical nuclear liquid gas transition and is partly due to temperature and partly to 
an attractive interaction between baryons. 
The binding energy per baryon in units of the baryon mass can at low temperatures be extracted from  
(here $N_f=1$ \cite{bind})
\beq
\epsilon= \frac{e-n_Bm_B}{n_Bm_B} =  -\frac{4}{3}\frac{1}{a^3n_B}\left(\frac{6h_1^3+3h_1^6}{z_0}\right)^2\,\kappa^2+\ldots\;.
\label{eq:bindpt}
\eeq
In the static limit this vanishes, for dynamical quarks it slowly grows with decreasing quark mass. This is also the 
reason why the onset in \fig\ref{fig:cont} (left) is a smooth crossover: $T_c\sim \epsilon$ is 
exponentially small for heavy quarks. For light quarks (larger $\kappa$) the expansion does not converge, but 
simulations clearly show a two-peak structure of baryon number at low temperatures, 
signalling a first-order onset transition, \fig\ref{fig:cont} (right).
Thus, all qualitative features of the nuclear liquid gas transition are contained in the effective theory and thus 
have been identified from QCD directly. 

\section{QCD at large $N_c$}

\begin{figure}[t]
\centering
\includegraphics[width=0.45\textwidth]{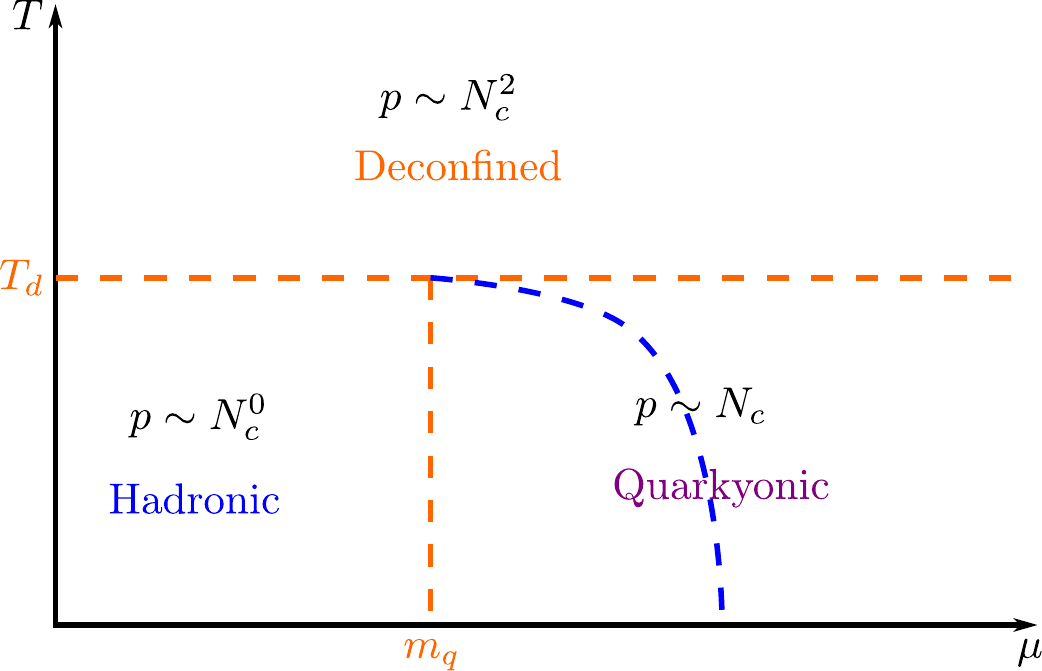}\hspace*{0.5cm}
\includegraphics[width=0.45\textwidth]{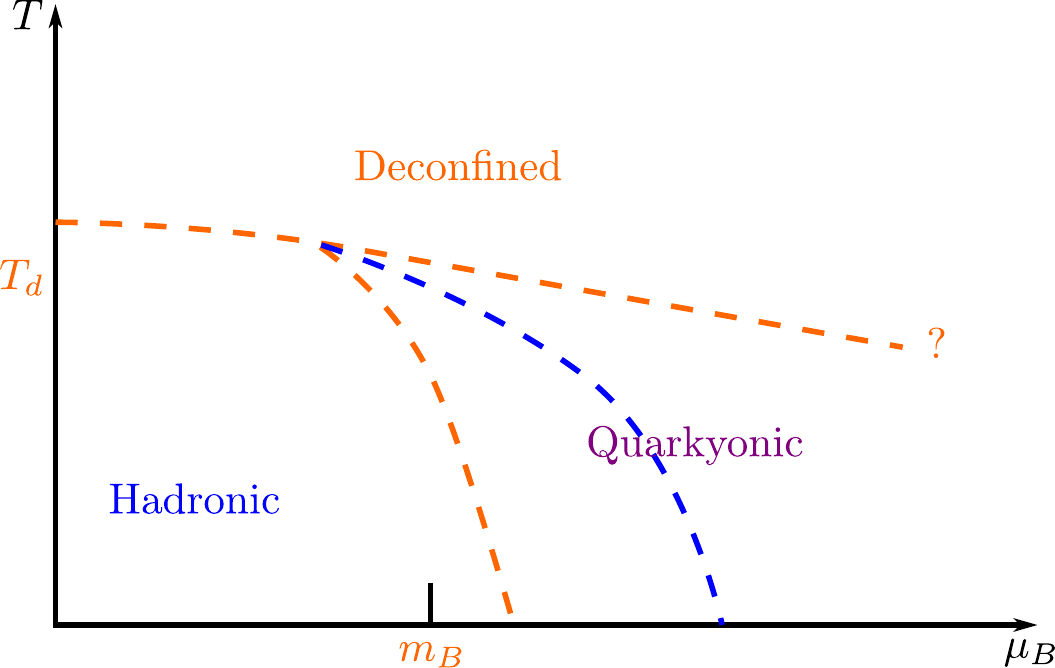}
\caption[]{Phase diagram in the limit of large $N_c$ (left) and possible consequences for $N_c=3$ (right) according to \cite{quarky}.
The blue line indicates the chiral transition. }
\label{fig:pd_nc}       
\end{figure}

Considering QCD with a large number of colours has a long history, with the initial hope to develop
an expansion scheme that also works for hadronic physics. Here we only   
summarise the most essential features established in the early works \cite{hooft,witten}.
The large $N_c$  limit of $SU(N_c)$-QCD is defined by 
\beq
N_c\rightarrow \infty\quad \mbox{with} \quad g^2N_c={\rm const.}
\label{eq:limit}
\eeq
In this case the theory has the following properties:
\begin{itemize}
\item Quark loops in Feynman diagrams are suppressed by $N_c^{-1}$
\item Non-planar Feynman diagrams are suppressed by $N_c^{-2}$
\item Meson masses are $\sim \Lambda_{QCD}$
\item Mesons are free; the leading corrections are cubic interactions $\sim N_c^{-1/2}$ and quartic interactions $\sim N_c^{-1}$
\item Baryons consist of $N_c$ quarks, baryon masses are $\sim N_c \Lambda_{QCD}$
\item Baryon interactions are $\sim N_c$
\end{itemize}

Using this, 
the authors of \cite{quarky} conjectured the phase diagram of QCD at large $N_c$ to look like \fig\ref{fig:pd_nc} (left). 
At large $N_c$, the influence of fermions on the deconfinement transition is suppressed, which therefore 
extends horizontally as a first-order transition into the chemical potential plane. In the deconfined phase, perturbation
theory is valid and the pressure scales as $p\sim N_c^2$. At low $T$ and $\mu$, the hadron resonance gas
is valid, the pressure is exponentially suppressed by hadron masses and scales as $p\sim N_c^0$. At large $\mu$ 
and low $T$ perturbation theory predicts quark matter scaling like $p\sim N_c$. The conjecture of \cite{quarky}
is that this regime reaches all the way down to $\mu\sim m_q$, where one expects baryon matter. 
The differences in scaling behaviour require non-analytic phase transitions between the three different regimes.
 Since matter in the cold and dense regime has both baryonic (towards its left boundary) and quark-like (at asymptotic densities) 
 features, it was termed ``quarkyonic''. A simple picture arises at low temperatures in momentum space, where
 fermions form a Fermi sphere, which constitutes the ground state. 
Excitations of order $\Lambda_\mathrm{QCD}$ relative to the Fermi sphere are then expected to be baryonic, whereas
excitations $\gg \Lambda_\mathrm{QCD}$ should be quark-like in nature. Thus in momentum space a shell structure
is expected \cite{quarky}, with an inner quark sphere surrounded by a baryonic shell of thickness 
$\sim \Lambda_\mathrm{QCD}$. The size of the entire and inner sphere is thus governed by $\mu$. 
A qualitative discussion of the $(T,\mu,N_c)$ phase diagram is also
given in \cite{mish}, and there are speculations about quarkyonic physics in heavy-ion collisions \cite{pbm} and
neutron stars \cite{reddy}. For an introduction, see \cite{mcl}.

\subsection{The liquid gas transition for general $N_c$}

\begin{table}[tbp]
\centering
\begin{tabular}{|c||c c c|}
\hline
 & $\kappa^0$ & $\kappa^2$ & $\kappa^4$ \\
\hline
$h_1<1$ &\multicolumn{3}{|c|}{}\\
\hline
$a^4p$ & $\sim \frac{1}{6 N_{\tau}} N_c^3 h_1^{N_c}$ & $\sim -\frac{1}{48} N_c^7 h_1^{2 N_c}$ & $\sim \frac{3 N_\tau \kappa^4}{800} N_c^8 h_1^{2 N_c}$ \\
     $a^3n_B$ & $\sim \frac{1}{6} N_c^3 h_1^{N_c}$ & $\sim -\frac{N_{\tau}}{24} N_c^7 h_1^{2 N_c}$ & $\sim \frac{(9 N_\tau+1)N_\tau}{1200} N_c^8 h_1^{2 N_c}$ \\     
     $\epsilon$ & 0 & $\sim -\frac{1}{4} N_c^3 h_1^{N_c}$ & \\
\hline
$h_1>1$ &\multicolumn{3}{|c|}{}\\
\hline
 $a^4p$ & $\sim \frac{4 \ln(h_1)}{N_{\tau}} N_c$ & $\sim -12 N_c$ & $\sim 198 N_c$ \\
 $a^3n_B$ & $\sim 4 $ & $\sim - N_{\tau} \frac{N_c^4}{h_1^{N_c}}$ & $ \sim -\frac{(59 N_\tau - 19) N_\tau}{20} \frac{N_c^5}{h_1^{N_c}}$ \\
 $\epsilon$ & 0 & $\sim -6$  & \\
\hline
\end{tabular}
\caption[]{Large $N_c$ behaviour of the thermodynamic functions and the interaction energy
per baryon, order by order in the hopping expansion, 
on both sides of the onset transition for $N_f=2$.}
\label{tab:scaling}
\end{table}

Within our analytic approach to lattice QCD, we can now employ the effective lattice theory to study the behaviour
of the baryon onset transition
as a function of increasing $N_c$ \cite{PS}. It should be stressed that we are \textit{not} doing any expansion in 
$N_c^{-1}$, so our results apply to any $N_c$ small and large. On the other hand, we do an expansion in the inverse
gauge coupling and quark mass, whose implications we will discuss below.

We begin by looking at the changes to the partition function in the static strong coupling limit, \eq(\ref{eq:zsc}), which now
reads
\beq
z_0=1+(N_c+1)h_1^{N_c}+h_1^{2N_c}\;.
\eeq
The modified prefactor indicates a different spin degeneracy of the baryon, which naturally depends on the number
of quarks that it is made of.

\begin{figure}[t]
\centering
\includegraphics[width=0.45\textwidth]{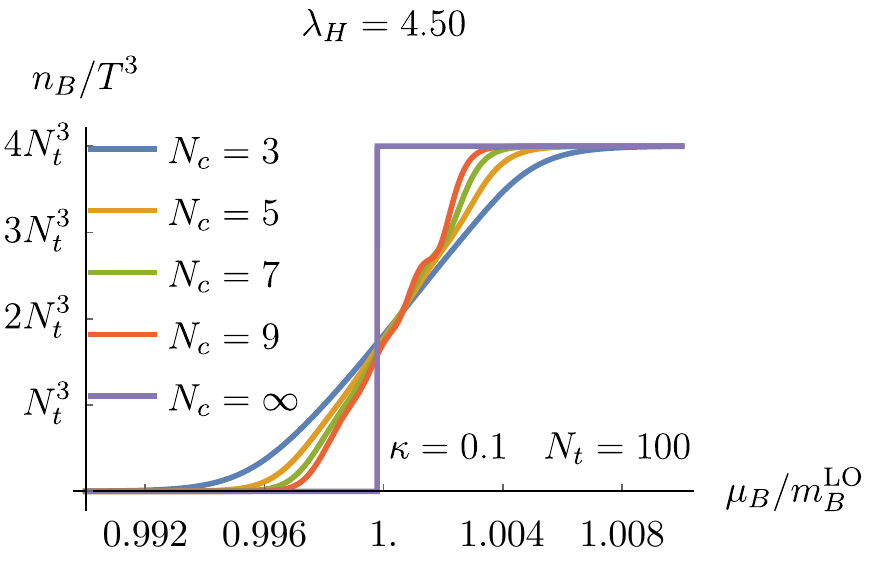}\hspace*{0.5cm}
\includegraphics[width=0.42\textwidth]{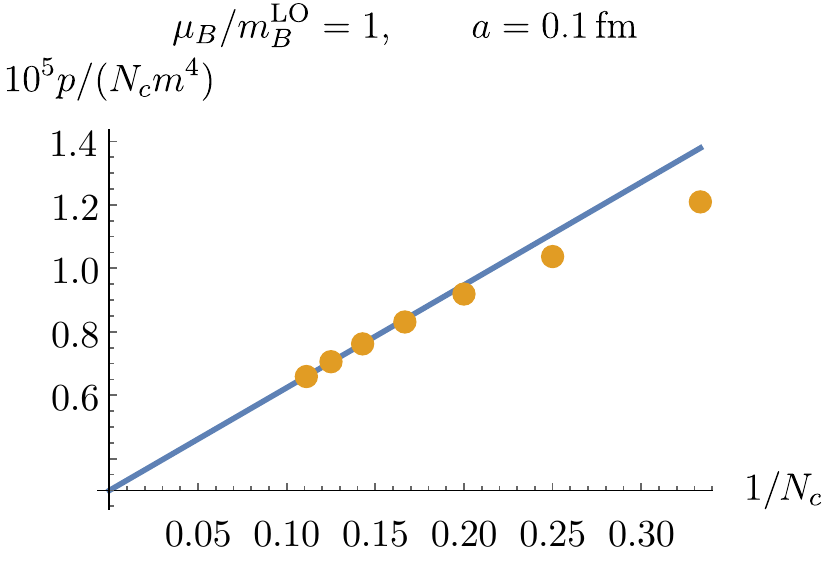}
\caption[]{Left: Onset transition for different values of $N_c$. Right: Pressure scaling as $p\sim N_c(1+{\rm const.\;}N_c^{-1}+\ldots)$ \cite{PS}.}
\label{fig:onset_nc}       
\end{figure}

Computing corrections in the hopping expansion, we obtain the coefficients for the pressure and baryon number
shown in Table \ref{tab:scaling}. Remarkably, for $h_1>1$, i.e., to the right of the onset transition, the coefficients of
all three computed orders are proportional to $N_c$, corresponding to part of the definition of quarkyonic matter.
For the LO coefficient this is trivial and given by the lattice saturation in terms of quark degrees of freedom.
However, the next two terms do not contribute to lattice saturation, but to the description of the physical baryon density which,
right at the onset transition, is composed of baryons, not quarks. The fact that these coefficients scale as $N_c$ is 
non-trivial and suggestive for this to be a feature to all orders. Note also that the binding energy per baryon is $N_c^0$
as expected \cite{quarky,witten}. 
On the other hand, for $h_1<1$ the contributions to the thermodynamic functions go as powers
of $h_1$, which vanish exponentially with $T\rightarrow 0$. Thus the silver blaze feature is amplified as $N_c$ increases,
while for $h_1>1$ the baryon density gets amplified with growing $N_c$. 
This steepens the onset transition with growing $N_c$, to ultimately always produce a first-order transition.

However, the results in the table were obtained in the strong
coupling limit $\beta=0$ and hence do \textit{not} yet correspond to the `t Hooft limit  \eq(\ref{eq:limit}), for which 
$g^2$ needs to be adjusted. Thus, including gauge corrections is mandatory. \fig\ref{fig:onset_nc} shows the
steepening of the onset transition and the pressure with gauge corrections included. It is remarkable
that the pressure scaling with $N_c$ needs only a leading correction to describe the behaviour almost 
down to $N_c=3$.
Finally, if we are interested in continuum 
physics, the ordering of limits matters, as already observed in \cite{gw}. The fact that for large $N_c$ the baryon density
jumps to the lattice saturation density, an artefact of the discretisation, indicates that we have
to take the continuum limit first and only then $N_c\rightarrow \infty$. Unfortunately, this prohibits a demonstration
of the large $N_c$ behaviour in the continuum, 
because an exploding number of orders in the expansions would be required to actually take the limits.
What is possible at present is to show that 
both qualitative features, 
i.e.~the steepening of the transition with $N_c$
and $p\sim N_c$ beyond onset, are stable when the lattice is made gradually finer
before increasing $N_c$ \cite{PS}.

\section{Conclusions}

\begin{figure}[t]
\centering
\includegraphics[width=0.42\textwidth]{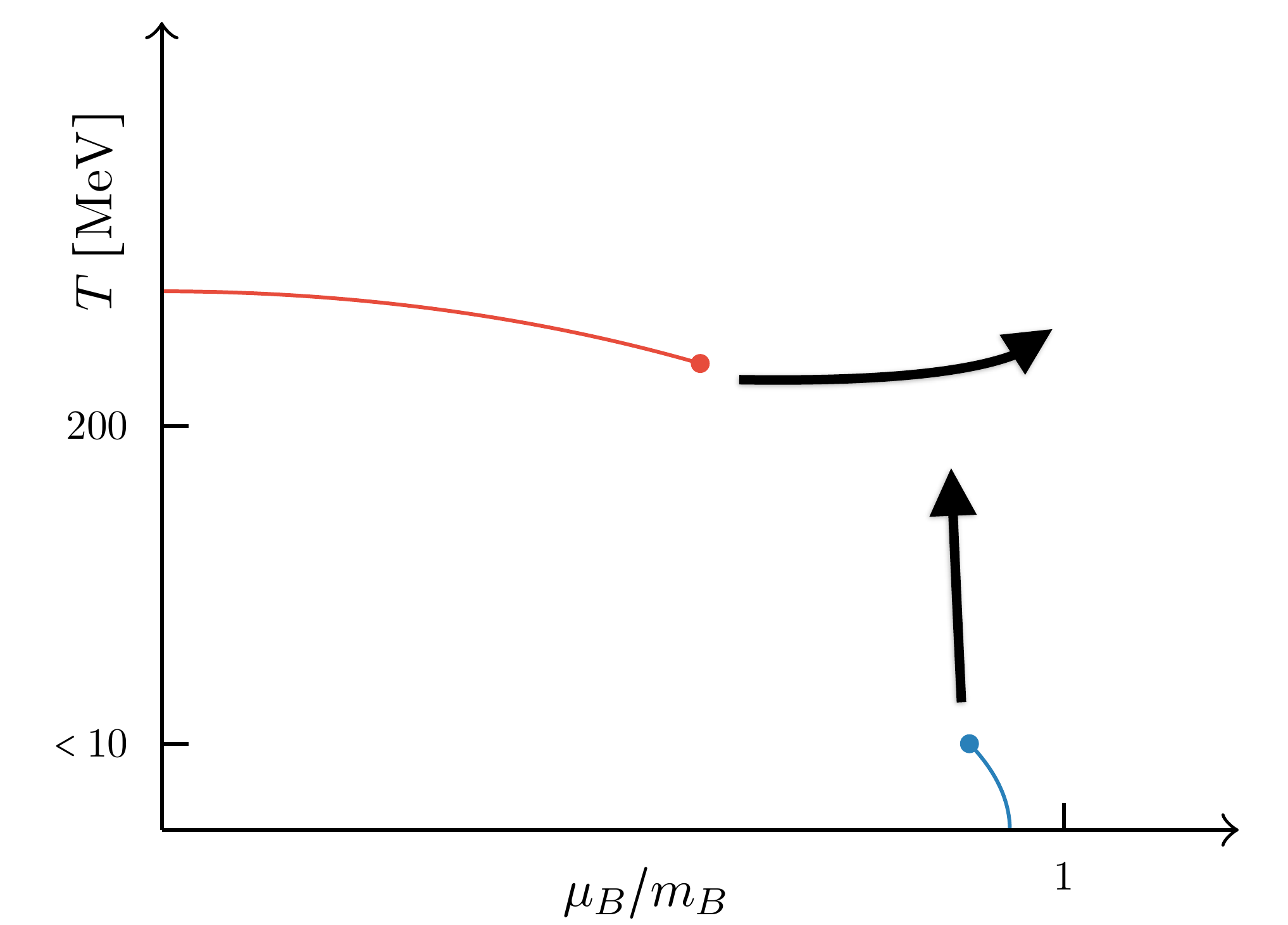}\hspace*{0.5cm}
\includegraphics[width=0.53\textwidth]{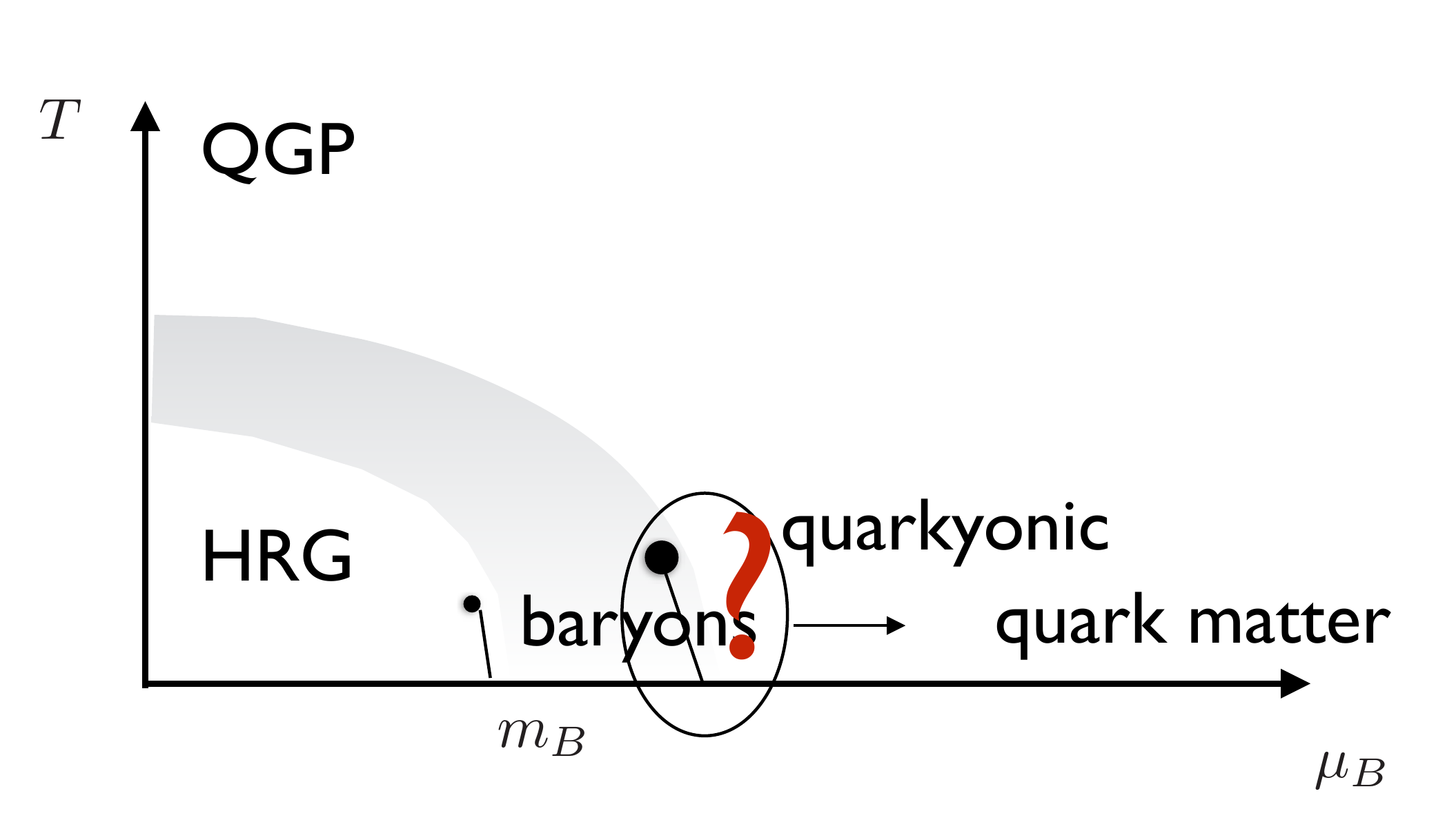}
\caption[]{Left: Smooth change of the transition lines for heavy QCD 
with growing $N_c$. Right: The features of the physical QCD phase diagram seen on the lattice.}
\label{fig:pd_phys}       
\end{figure}

We have studied the onset transition to baryon matter for general number of colours $N_c$ with an 
effective lattice theory for heay quarks, which is derived from full lattice QCD 
by analytic character and hopping expansion methods.
The onset transition becomes
more strongly first-order with increasing $N_c$, which implies increasing $T_c$ for its endpoint as well, 
cf.~\fig\ref{fig:pd_phys} (left). 
Since the deconfinement transition is known to ``straighten'' with growing $N_c$, 
we observe how the phase diagram of QCD with heavy quarks gradually evolves towards the conjectured rectangular shape
in the large $N_c$ limit, \fig\ref{fig:pd_nc} (left). We also confirm the pressure beyond the onset transition 
to scale as $p\sim N_c$ in that limit. If this finding generalises to all orders in the hopping expansion, then it also holds for
light quarks and cold and dense QCD is consistent with quarkyonic matter as formally defined in \cite{quarky}.

What are the consequences for physical QCD? For $N_c=3$, there is no separate phase transition
to quarkyonic matter as in \fig{\ref{fig:pd_nc} (right), that role is played by the nuclear liquid gas transition,
which terminates at $T_c\sim 16$ MeV.
There is so far no lattice evidence for the other transition lines either, which might be crossovers.
Right after onset there are then baryons only, which is consistent with the shell picture of quakyonic matter and similar  
to the finite temperature 
crossover region still being mostly hadronic \cite{gloz}. This is indicated by the shaded area in \fig{\ref{fig:pd_phys} (right). 
With increasing density one would then expect the inner sphere of quark matter to form, allowing for a possibly 
smooth transition to predominantly quark matter at large densities, 
similar to the smooth transition to quark gluon plasma in the temperature direction. A separate study with light
quarks is needed to see whether or not there is in addition a chiral transition.

\noindent
{\bf Acknowledgments}: 
The authors acknowledge support by the Deutsche Forschungsgemeinschaft (DFG) through the grant CRC-TR 211 ``Strong-interaction matter
under extreme conditions''.

\end{document}